# Suppression of a charge density wave ground state in high magnetic fields: spin and orbital mechanisms

D. Graf, J.S. Brooks, and E.S. Choi

NHMFL/Physics, Florida State University, Tallahassee, FL 32310, USA

S. Uji

National Institute for Materials Science, Tsukuba, Ibaraki 305-0003, Japan

J.C. Dias and M. Almeida

Dept. de Quimica, Instituto-Tecnologico e Nuclear, P-2686-953 Sacavem, Portugal

M. Matos

Dept. de Engenharia Quimica; Instituo Superiore de Engenharia de Lisboa, P-1900

Lisboa, Portugal

The charge density wave (CDW) transition temperature in the quasi-one dimensional (Q1D) organic material of $(Per)_2Au(mnt)_2$ is relatively low ($T_{CDW}$ = 12 K). Hence in a mean field BCS model, the CDW state should be completely suppressed in magnetic fields of order 30 – 40 T. To explore this possibility, the magnetoresistance of $(Per)_2Au(mnt)_2$ was investigated in magnetic fields to 45 T for 0.5 K < T < 12 K. For fields directed along the Q1D molecular stacking direction, $T_{CDW}$ decreases with field, terminating at about ~ 37 T for temperatures approaching zero. Results for this field orientation are in general agreement with theoretical predictions, including the field dependence of the magnetoresistance and the energy gap, $\Delta_{CDW}$. However, for fields tilted away from the stacking direction, orbital effects arise above 15 T that may be related to the return of un-nested Fermi surface sections that develop as the CDW state is suppressed. These findings are consistent with expectations that quasi-one dimensional metallic behavior will return outside the CDW phase boundary.

# Introduction

The effects of magnetic fields on the charge density wave ground state have been a long-standing area of interest. In general, for increasing fields, the Zeeman splitting of the bands at the Fermi level will reduce the pairing interaction. This eventually leads to a non-condensate, metallic state where the energy gap is driven to zero. The similarity between the BCS and CDW ground states was considered by Dieterich and Fulde [1], who predicted a field dependence of the CDW transition

$$\frac{\Delta T_{CDW}}{T_{CDW}(0)} = -\frac{\gamma}{4}\left(\frac{\mu_B B}{k T_{CDW}(0)}\right)^2 \qquad (1)$$

where the relevant variable is the $\frac{B}{T_{CDW}}$ ratio. The pre-factor $\gamma$ generally is of order 1, and will depend on spin-orbit interactions in tilted magnetic fields away from the chain direction. This theory was extended to predict the nature of the magnetoresistance (MR) in the CDW state by Tiedje and co-workers [2] who showed that for T << $T_{CDW}$,

$$\frac{\Delta \rho}{\rho} = -\frac{1}{2}\left(\frac{\mu_B B}{kT}\right)^2 + 0\left(\frac{\mu_B B}{kT}\right)^4. \qquad (2)$$

The main effect of the field is to increase the number of carriers as the spin-up band moves down and the gap decreases. For T >> $T_{CDW}$, outside the range of fluctuations, there should be no MR for a 1D system in the normal metallic state. It was also shown in Ref. [2] that the magnetic field dependence of the CDW gap $\Delta_{CDW}$ should follow the BCS-like behavior. These predictions were tested to 5 T on the CDW system TTF-TCNQ, but since the CDW transition temperature was relatively high (52 K ), $\frac{\mu_B B}{k T_{CDW}}$ was only about 0.1, and the changes in the MR were limited to a few percent.

More recently, a class of organic metals, (Per)$_2$M(mnt)$_2$ (perylene – metal – maleonitriledithiolate: M = Au, Pt, Pd, Ni, Cu, Co, Fe ) [3] have been shown to be CDW systems, based on diffuse x-ray studies[4] and non-linear transport measurements [5,6]. These highly anisotropic materials consist of nearly isolated chains of perylene molecules along the most conducting b-axis. Hence the Fermi surface should be, to a good approximation, two parallel sheets with Fermi momentum in the b-axis direction. For M

= Au, there is no magnetic moment in the M(mnt)$_2$ anion structure, and hence this system is close to an ideal Peierls system. The crystal structure of (Per)$_2$M(mnt)$_2$ is presented in figure 1. The conducting chains of perylene alternate in the a,c plane with chains of anions in such a way that each stack of anions is surrounded by six stacks of perylene and each perylene stack has 3 stacks of perylene and 3 stacks of anions as nearest neighbors [7]. The alternating packing pattern of donor and acceptor chains is such that there does not appear to be any well defined quasi-two dimensional (Q2D) "conducting planes" as there is, for instance, in the Q1D Bechgaard salts or in α-(BEDT-TTF)$_2$KHg(SCN)$_4$ which exhibits both closed orbit oscillations and density wave transitions [8, 9]. The presence of hydrogen atoms in the periphery of the perylene molecules, which have no contribution to the highest occupied molecular orbital (HOMO), render the electronic interactions mediated by the interchain contacts ineffective and the system is expected to behave as almost purely one dimensional.

Recent band structure calculations show the electronic anisotropy of Per$_2$Au(mnt)$_2$ to be approximately 750:10:1 for $t_b$:$t_a$:$t_c$ respectively [10]. For the perylene chains, the intrachain band width $t_b$ is of order 149 meV, the interchain bandwidth in the a-axis direction $t_a$ is of order 2 meV, and the interchain bandwidth in the c-axis direction $t_c$ is of order 0.2 meV or less. Compared with other Q1D systems, the largest a-axis inter-chain bandwidth is an order of magnitude less than it is in standard Q1D metals such as the Bechgaard salts[8] (~ 22 meV).

Of major relevance to the work presented here is that $T_{CDW}$ in these materials approaches the lowest value yet observed for a CDW transition (for M = Au, $T_{CDW}$ ~ 12 K; M = Pt, $T_{CDW}$ ~ 8 K), and with contemporary high magnetic field facilities, the range $\frac{\mu_B B}{kT_{CDW}} > 1$ is accessible. Our work follows previous work by Matos *et al.* to 18 T, where $T_{CDW}$ was suppressed by about 15% and 40% in (Per)$_2$M(mnt)$_2$ for M = Au and M = Pt, respectively [11]. The purpose of the present work, therefore, has been to test our present understanding of the effects of high magnetic fields on a CDW system for the full range of magnetic fields and temperature within which the CDW state is bound. We find that for a magnetic field parallel to the 1D stacking direction, the results closely follow the theoretical predictions, with an upper critical field above 30 T for temperatures approaching zero. However, for field components perpendicular to the stacking direction, we find that the suppression of the CDW is anisotropic, and at high fields evidence for an un-gapped Q1D Fermi surface appear.

**Experimental**

The $(Per)_2M(mnt)_2$ materials used in this investigation were grown using electrocrystallization techniques described previously[12]. For each sample, the ac MR was measured along the b-axis using a standard four-terminal configuration of 10 $\mu$m Au wires attached with carbon paste. Two samples were measured in dc magnetic fields to 33 T (hereafter $S_1$ and $S_2$), and four other samples were measured in higher fields to 42 T ($S_3$-$S_6$). Samples 1 through 5 were placed on a rotator probe in a $^3$He cryostat system, where the b-axis of the samples, and also the field direction, were normal to the rotation axis. In the present notation, for $\theta = 0^0$, the field was along the b-axis (chain) direction (B//b), and for $\theta = 90^0$ the field was perpendicular to the b-axis (B$\perp$b). Sample $S_6$ was placed with the b-axis parallel to the axis of rotation, so that the field was tilted in the ac-plane, with the field perpendicular to the b-axis. The sizes of the needle crystals were of order 2 x 0.05 x 0.02 mm$^3$ where the long axis is the crystallographic b-axis.

**Results**

The magnetic field dependence of the MR of four samples of $(Per)_2Au(mnt)_2$ are shown in Fig. 2 as a function on field direction at 0.5 K. (The data was taken in the NHMFL hybrid magnet, where the superconducting outsert is maintained at 11.5 T. Hence, lower fields were not accessible during these experiments.) For measurements with the applied field parallel to the chain axis ($\theta = 0^0$) the resistance is observed to decrease monotonically by several orders of magnitude, and approaches an asymptotic value above 32 T. We have defined $B_{CDW}$ as the intersection of the two slopes in the MR data (decreasing and asymptotic) as shown in the figure. Above $B_{CDW}$, the MR for B//b appears to vanish in a manner consistent with a purely 1D conductor. As the field is tilted towards the $B \perp b$ direction (Figs. 2a-2c), $B_{CDW}$ increases, and the asymptotic resistance also increases. In all cases, in tilted fields additional structure appears in the MR (~20-30T), and at the highest fields the resistance shows an upturn above 35 T. The overall behavior of the field dependence of the first three samples in Fig. 2, including the high field angular dependence of the asymptotic resistance (i.e. and upturn above 35 T), is similar. In the case where the field was rotated in the ac-plane (Fig. 2d for $B \perp b$), the MR continues to be highly anisotropic. With the field applied parallel to the a-axis $\Delta\rho/\rho$ follows similar behavior to the B // b-axis orientation. When the field is rotated near B // c-axis orbital effects raise the high field MR approximately 50%.

The activated behavior of the conductance (taken as the inverse of the b-axis resistance) vs. inverse temperature for different field values and field orientations is shown in Fig. 3a and 3b for sample $S_4$, and in Fig. 3c for sample $S_5$. We note that in the low temperature limit, the sample conductance deviates from the expected $\sigma_0 \exp(-\frac{\Delta}{2kT})$ activated behavior. Several competing factors may be involved including Joule heating, de-pinning of the CDW, and residual conductivity due to un-nested Fermi surface sections. In our constant-current bias measurement, the voltage (and electric field) will rise as the as the sample resistance rises for decreasing temperature below $T_{CDW}$. For the highest resistance data (B ≤ 25 T) for sample $S_4$ where 10 $\mu$A was used, we have applied a simple linear Joule-heating model ( $IV = k_{eff}(T_s - T_b)$) to account for the difference in the sample temperature ($T_s$) with respect to the helium bath temperature ($T_b$). The effective thermal conductivity ( $k_{eff} \approx 3\frac{\mu W}{K}$ ) is between the sample and the $^3$He exchange gas. The conductance plotted against the computed sample temperature $T_s$ is shown for the low field (B ≤ 25 T) data (dashed curves), and the Arrhenius behavior is restored. Only for the highest resistance (lowest conductance) data of the 11.5 T temperature curve does the electric field across the sample below 3 K exceed the threshold field for de-pinning of the CDW, as previously reported by Lopes *et al.* [5,6] for the same material. There appears to be some evidence for this in the lowest temperature part of the conductance data. In Fig. 3c for $S_5$, the lower excitation current (1 $\mu$A) did not produce Joule heating. For the data at higher fields (B > 25 T) in Figs. 3a and 3b, we were not able to account for the excess conductance by the Joule heating model. Likewise, the electric fields in this field range were well below (factors of $10^{-2}$ or less) the threshold electric fields for the B = 0 case. Above $B_{CDW}$, the conductivity still appears slightly activated, especially for the B$\perp$b sample orientation. We believe the excess conductivity, and its temperature and field dependence, arises from the restoration of the Fermi surface at higher fields as the CDW is destroyed and possibly from transitions to a new CDW-type state as discussed below.

We have determined the thermal activation energy for each field (Figure 4a) in the temperature range below $T_{CDW}$ by Arrhenius ($\sigma = \sigma_0 \exp(-\frac{\Delta}{2kT})$) analysis. The curves were fit below $T_{CDW}$ (heavy dotted lines), when the gap was well-developed but before the sample resistance begins to differ from the exponential behavior. The energy gap (Following Bonfait et al [13], we have used the logarithmic derivative method, as shown in figure 4b, to determine the values of $T_{CDW}(B)$. The peaks are easily observed to fields of 29 tesla but could not be resolved for higher fields.

A summary of the field dependence of $T_{CDW}(B)$ and $\Delta_{CDW}(B)$ for (Per)$_2$Au(mnt)$_2$ is given in Fig. 5. For B//b the behavior of $T_{CDW}$ closely follows the $\frac{\gamma}{4}(\frac{\mu_B B}{kT_{CDW}(0)})^2$ dependence discussed above, where $\gamma \approx 1$ is expected for a conventional Peierls transition[14]. At lower fields (B < 25 T) the results seem to be angle independent and with a γ value in agreement with that found previously by Matos and Bonfait et al. [11, 15, 16]. At high field and at low temperature, as shown also in Fig 2a-2d for T=0.45 K, rather strong anisotropic effects occur. The energy gap $\Delta_{CDW}$ calculated in the range 3-10 K for sample S$_2$, decreases with increasing field in accord with a BCS-like dependence, as expected from the mean field theory [2]. One further test of the mean field theory is the field and temperature dependence of the MR discussed above. We find that the MR data for B//b follows an approximate $(\frac{B}{T})^2$ dependence, but at high fields near $B_{CDW}$ the dependence on B and T is more complicated and a comparison with the mean field theory is not possible.

## Discussion

The most basic description of our results follows from the picture that the magnetic field destroys the CDW state as the Pauli spin energy exceeds the CDW condensation energy. For a truly 1D system, in the absence of orbital coupling, $T_{CDW}(B)$ will be isotropic with field direction. During this process, un-nested parts of the Fermi surface will eventually re-appear at high fields. When the field is directed along the b-axis, parallel to the perylene chain stacking axis, it cannot couple orbitally to any Q1D Fermi surface sections that emerge, and hence the MR will drop to the metallic value at $B_{CDW}$ as indicated clearly in the three samples shown in Fig. 2a,b, and c for B//b ($\theta = 0^0$). The field dependence of $T_{CDW}$, even in the case where $\theta \neq 0^0$, still follows the mean field BCS like description of the CDW state (although $B_{CDW}$ increases and orbital effects start to appear). When the field is perpendicular to the conducting chains ($\theta = 90^0$), mechanisms involving orbital effects have a maximum effect. In Fig. 2d, we find that for field perpendicular to the a-b plane (i.e. B//c), the MR is largest. This is consistent with a small, but significant interchain bandwidth in the a-axis direction. Although the assignment of the structure seen in the MR data in Fig. 2 for $\theta \neq 0^0$ is not possible at this stage, it is clear that orbital effects are present. The structure in the MR in all cases just below $B_{CDW}$ moves to higher fields with increasing angle, and the upturn in the MR above 35 T also seems to be a general feature for $\theta \neq 0^0$. For the case where the field was rotated in the ac plane (Fig.2d), as discussed above, we obtain the largest angular dependence of the MR between 0 and 90 degrees.

This suggests that not only are there orbital effects, but that they are anisotropic in the ac plane, and that for one orientation B//c, there may be a larger inter-chain coupling which leads to a larger MR than for B//a. This is consistent with the anisotropy between $t_a$ and $t_c$ from the band calculations[10]. Given the differences in MR structures in the different samples, and their relatively low frequency (estimated for a single period to be between

20 and 200 T) compared with standard quantum oscillations in Q2D organic metals (typically above 200 T), the inter-chain coupling may involve very small changes in the crystallographic structure at low temperatures which drive the corresponding electronic structure away from purely 1D.

An interesting possibility is that the magnetic field may actually induce orbital (spin density wave or SDW) nesting [8] in the Q1D Fermi surface as the CDW ground state is removed. Fig. 3 gives some suggestive evidence that field induced nesting may occur, since we find that even well above $B_{CDW}$, the conductivity is still strongly activated for $B \perp b$, ie. $\theta = 90^0$. In Fig. 3c for the 42 tesla data ($\theta = 90^0$) the activation energy is $\Delta/2 =$ 8K. Hence, although the nearly isotropic CDW gap is closed, there is evidence that an orbitally induced gap opens at higher fields.

Recent theoretical work on the magnetic field dependence of a CDW ground state[17, 18], beyond the simple mean field treatment in Eq.1, is potentially relevant to the behavior of $Per_2Au(mnt)_2$. The theory involves a quasi-one-dimensional CDW ground state, where both spin and orbital terms are included in an anisotropic, two-dimensional Hubbard model. The model predicts a transition from the low field ground state $CDW_0$, through a transition, to a high field ground state $CDW_x$ and/or $CDW_y$ depending on the ratio of the spin and orbital coupling and on the direction of the applied magnetic field. The $CDW_x$ order is exclusively from the Pauli effect while the $CDW_y$ order from both the Pauli and orbital effect. Recent work on the α-$(ET)_2KHg(SCN)_4$ systems has indicated a remarkable correspondence between the high field ground states and the theoretical predictions[9, 19-21], in spite of the Q2D nature of the electronic structure.

For finite field, especially perpendicular to the a-b plane where, as mentioned above, orbital effects might be expected, the data of Fig. 2 suggests some correspondence with theory. Here, below 35 T, the state is CDW, there is a low resistance minimum, and at higher fields the resistance rises again. The CDW (for instance, $CDW_0$) resistance appears to be activated. The high field behavior above 35 T is weakly activated, but even higher fields than 45 T may be necessary to determine if a second, gapped state (for instance, $CDW_{x(y)}$) is stabilized. Nonetheless, some experimental observations may be relevant with the transition to a new CDW orders as the theory predicts. The Arrhenius plots shown in Fig. 3 show change of slopes even well below the threshold electric field, which could not be explained by the Joule heating model. This may be due to the 2nd order transition from the $CDW_0$ to $CDW_x$ (for B || b) or to $CDW_y$ (for B is tilted from b). The increase of $B_{CDW}$ by tilting magnetic field away form the b-axis is also predicted by the theory, in which $B_{CDW}$ corresponds to $h_{cy}$ in Ref. [18].

Before we conclude, it should be noted that the above theory was based on the perfect nesting case and the $CDW_x$ order does not depend on $t_a$ while $CDW_y$ requires a finite $t_a$. In the case of $Per_2Au(mnt)_2$, for the interchain energy $t_a$, the relevant parameter in the theory is the parameter $t_a'/t_a'^*$ where $t_a'$ is the imperfect nesting parameter[8] (of order $t_a^2/t_b$), and $t_a'^*$ is related to the perfect nesting transition temperature $T^0_{cdw}(t_a'=0)$ (= 1 meV in the present case). Hence $t_a'/t_a'^* = 0.03$, and this puts the $Per_2Au(mnt)_2$ system in the limit of nearly perfect nesting, with respect to the theory. Hence the structure in $T_{cdw}(B)$ predicted by the theory for less perfect nesting in the $CDW_0$ state should be suppressed in the present case. However, some kind of structure appears in Fig. 2, within the $CDW_0$ state. Furthermore, finite values of $t_a'$ were suggested to explain the discrepancy between the values of pre-factor $\gamma$ in Eq. (1) and experimental results of Matos et al. (and also results of this work) [18]. The $t_a'$ value needed to explain the discrepancy was estimated to be about 7.4 K, which corresponds 750:50 for $t_b$:$t_a$ ratio, compared with the ratio 750:10 from band structure. It is possible that, although surprising, the low temperature inter-chain band width may be significantly larger that that expected from the band calculations.

## Conclusions

The main results of the present work are: i) for the non-magnetic anion member (M = Au) of the $Per_2M[mnt]_2$ class of highly one-dimensional materials, a magnetic field suppresses the CDW ground state above ~ 37 T, when the field is parallel (B//b) to the conducting chains; ii) for finite field in the a-c plane, orbital effects arise that give a more complicated dependence of the magnetoresistance both below and above the field at which the CDW-to-metal phase boundary is expected at low temperatures. This last result implies that even in this highly one-dimensional system, inter-chain, orbital effects exist,, and that the inter-chain bandwidth is larger than expected from band calculations. iii) At low fields, mean field theory is in good agreement with the suppression of a conventional CDW state, but at high fields, a theoretical treatment that includes both spin and orbital coupling to the magnetic field appears to be relevant. There is strong evidence for orbital mechanisms in high fields that interact with an anisotropic, quasi-one-dimensional Fermi surface, which may induce a new nesting and/or quantum oscillatory behavior. Hydrostatic and uniaxial pressure studies, as well as experiments to even higher magnetic fields would be useful to fully explore these possibilities.

## Acknowledgements

This work is supported by NSF-DMR 02-03532, and the NHMFL is supported by a contractual agreement between the NSF and the State of Florida. DG is supported through a NSF GK-12 Fellowship. Work in Portugal was supported by FCT under contract POCT/FAT/39115/2001. Helpful discussions with R. T. Henriques and G. Bonfait are also acknowledged.

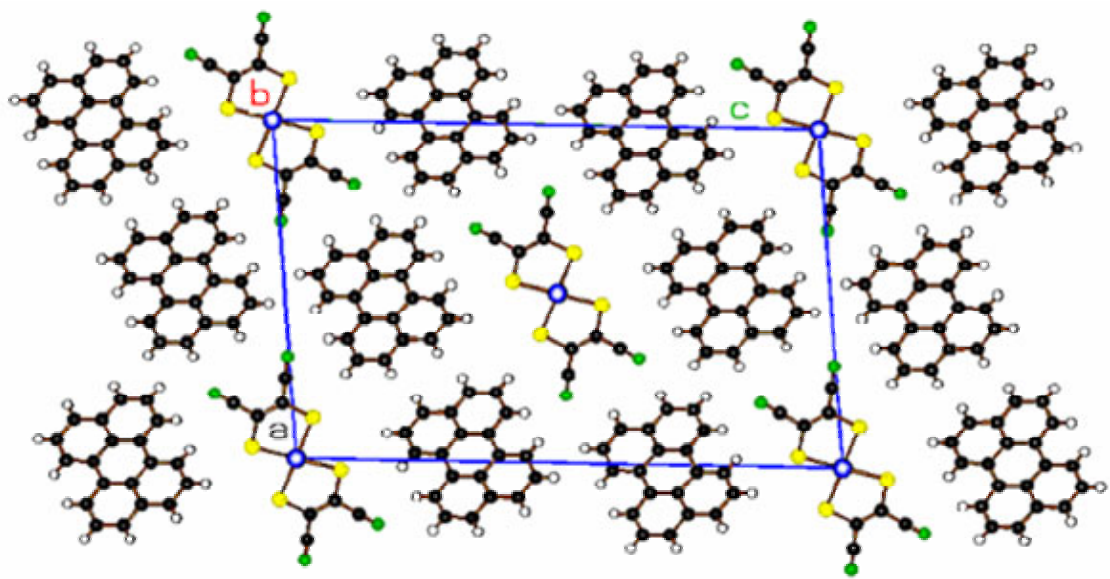

Figure 1. Crystallographic projection of (Per)$_2$M(mnt)$_2$ along the perylene b-axis stacking direction [Ref. 9].

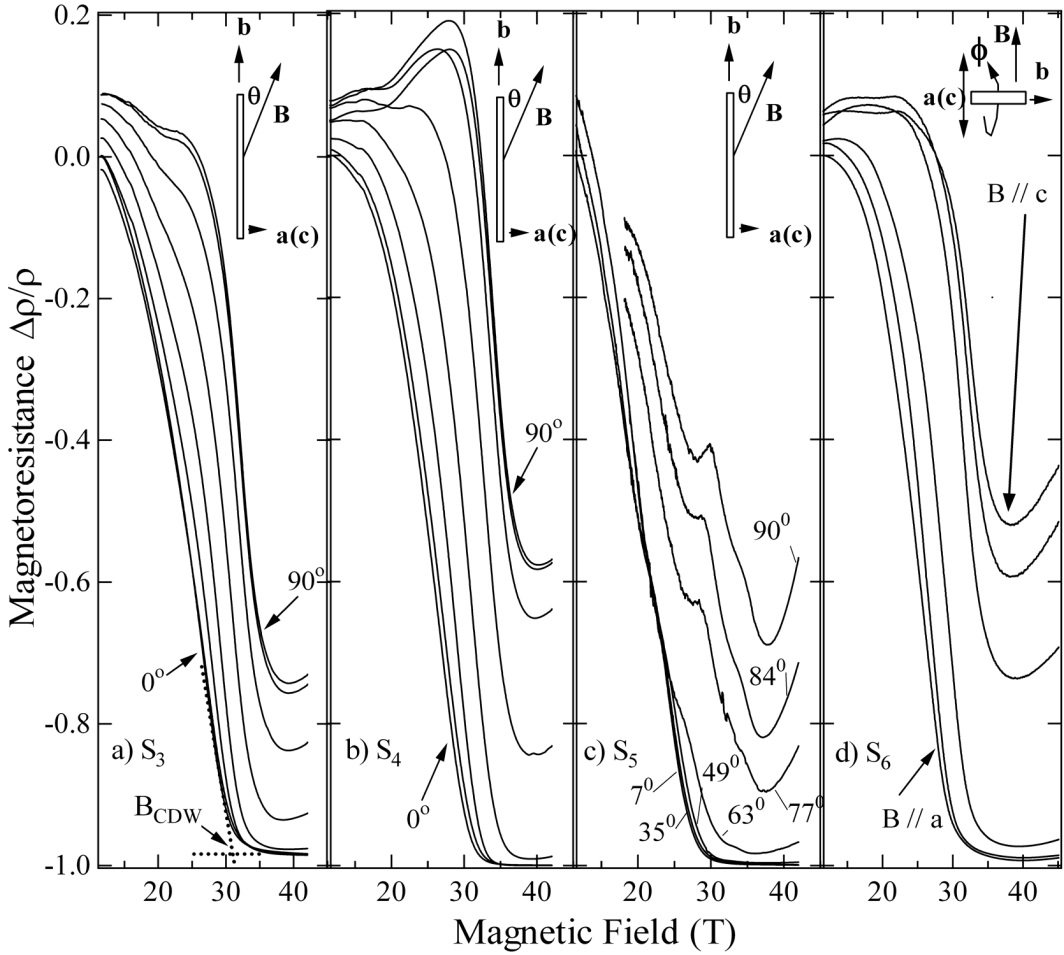

Figure 2. Magnetic field orientation dependence of the MR of (Per)$_2$Au(mnt)$_2$ at 0.45 K for different samples and sample configurations. The MR data is normalized to the B = 11.5 T and $\theta = 0°$ orientation values. Figs. 2a), 2b), and 2c): polar ($\theta$) rotation of field B with respect to the b-axis in different undetermined azimuthal ($\phi$) b-a(c) planes. Fig 2d): Azimuthal ($\phi$) rotation in the a(c) plane for field B perpendicular to the b-axis. Note that the MR is largest for the B//c orientation (see text). The upper critical field for the CDW-metal transition $B_{CDW}$ is defined by the intersection of the CDW and metallic MR slopes as shown in 1a. Angles not specifically noted are as follows: Figs. 2a and 2b) $\theta$ = 0, 13,

27,41,55,69, 83, 90 degrees, Fig. 2d) $\phi$ = 0, 17.5, 35, 52.5, 70, 87.5 degrees from arbitrary initial $\phi$ (azimuthal rotation).

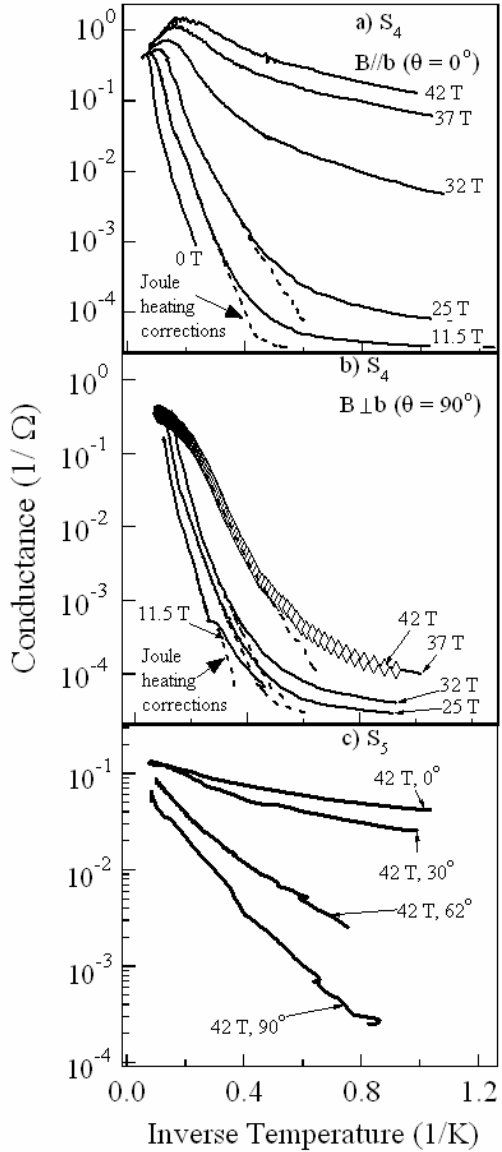

Figure 3. Arrhenius plots of conductance (1/R) of $(Per)_2Au(mnt)_2$ vs. inverse temperature for variable magnetic field values and sample orientations. Dashed lines are for the high resistance (low conductance - less than $10^{-3}$ $1/\Omega$) data corrected for Joule heating effects. a) Sample $S_4$: B//b (field parallel to the staking axis). b) Sample $S_4$: B$\perp$b (field

perpendicular to the stacking axis). c) Sample $S_5$: Data at 42 T vs. angle where $0^0$ indicates B//b and $90^0 = B \perp b$.

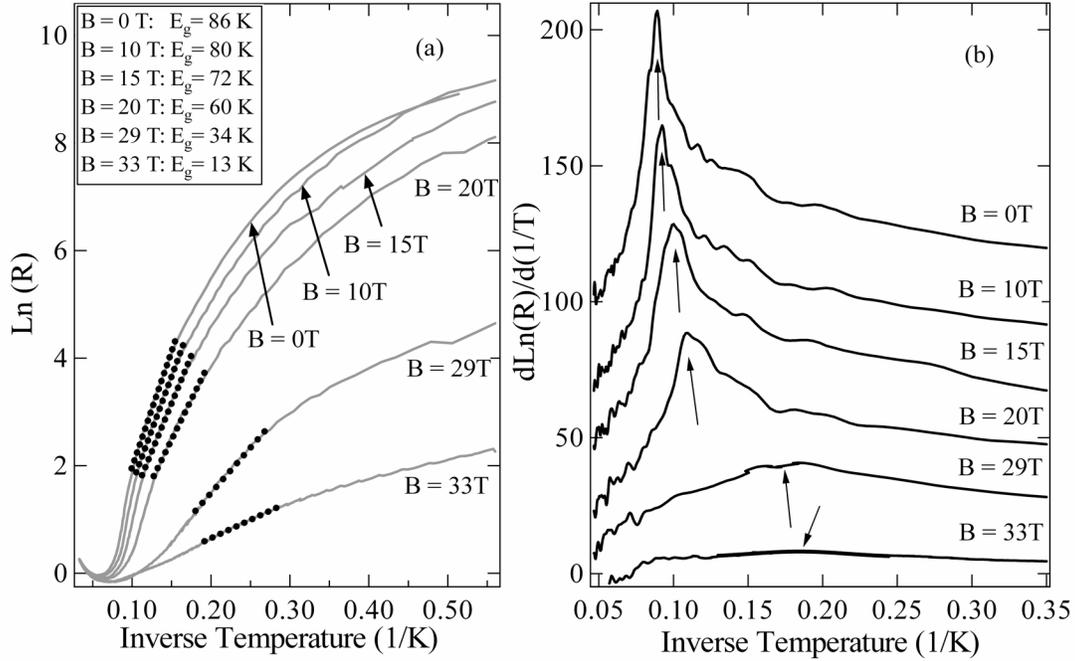

Figure 4. (a) Arrhenius plot of the temperature dependence of the resistance vs. field for sample $S_2$. The ranges of the fits used to obtain the field dependent activation energies are shown by the dotted lines. (b) Logarithmic derivatives of the data in (a) used to obtain the field dependent values of $T_{CDW}$ (arrows).

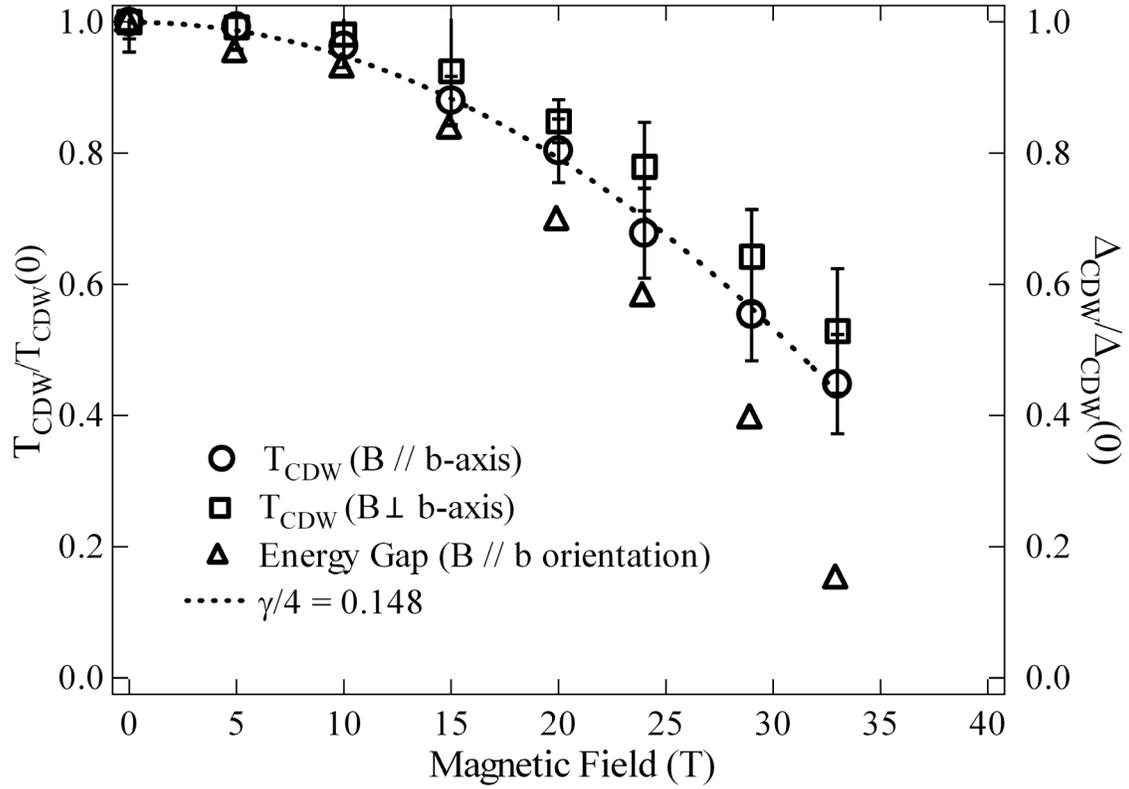

Figure 5. Summary of $T_{CDW}(B)/T_{CDW}(0)$ (for parallel and perpendicular sample orientations) and $\Delta_{CDW}(B)/\Delta_{CDW}(0)$ (for B parallel to the b-axis). $T_{CDW}(0) \approx 11.3$ K for all samples and $\Delta_{CDW}(0) = 88$ K (sample $S_2$). The dotted line shows the theoretical fit for equation (1) with a pre-factor of $\gamma/4 = 0.148$ (in agreement with the results of Ref. 11).